\documentclass[12pt]{article}
\usepackage{epsf}
\usepackage{amsfonts} 
\usepackage{amsbsy} 

\newdimen\hsgraph \newdimen\vsgraph
\hsgraph=\hsize \advance\hsgraph by-1.2truein
\vsgraph=\vsize \advance\vsgraph by-1.5truein

\newcommand{\C}[1]{{\mathcal #1}}

\newcommand{\R}[1]{{\mathrm #1}}

\newcommand{\beq}[1]{\begin{equation}{\label{#1}}}
\newcommand{\eeq}{\end{equation}}
\newcommand{\bea}{\begin{eqnarray}}
\newcommand{\eea}{\end{eqnarray}}
\newcommand{\rf}[1]{(\ref{#1})} 
\newcommand{\nn}{\nonumber}

\newcommand{\half}{{1\over 2}}
\newcommand{\twothirds}{{2\over 3}}

\newcommand{\fourthirds}{{4\over 3}}

\newcommand{\gstr}{\gamma}

\newcommand{\zcrit}{z_{\R c\R r}}
\newcommand{\Qt}{{\widetilde{\C Q}}}
\newcommand{\yy}{{\Bigg |_{y=1}}}
\newcommand{\union}{\cup}
\newcommand{\unionsum}{\sum_{\R A_1\union\R A_2\union\ldots\R A_n=\R A}
z^{1+N_{{\R A}_1}+\ldots N_{{\R A}_n}}w_{{\R A}_1}\ldots w_{{\R A}_n}}

\begin{document}
\topmargin 0pt
\oddsidemargin 5mm
\headheight 0pt
\topskip 0mm

\addtolength{\baselineskip}{0.20\baselineskip}

\pagestyle{empty}

\begin{flushright}
OUTP-97-63P\\
27th November 1997\\
hep-th/9712058
\end{flushright}

\begin{center}

\vspace{18pt}
{\Large \bf The Spectral Dimension of Non-generic Branched Polymer
Ensembles}

\vspace{2 truecm}

{\sc Jo\~ao D. Correia\footnote{e-mail: j.correia1@physics.ox.ac.uk} 
and John F. Wheater\footnote{e-mail: j.wheater1@physics.ox.ac.uk}}

\vspace{1 truecm}

{\em Department of Physics, University of Oxford \\
Theoretical Physics,\\
1 Keble Road,\\
 Oxford OX1 3NP, UK\\}

\vspace{3 truecm}

\end{center}

\noindent
{\bf Abstract.} We show that the spectral dimension on non-generic
branched polymer models with susceptibility exponent
 $\gstr>0$ is given by $d_s=2/(1+\gstr)$.
For those models with $\gstr<0$ we find that $d_s=2$.

\vfill
\begin{flushleft}
\end{flushleft}
\newpage
\setcounter{page}{1}
\pagestyle{plain}

The spectral dimension $d_s$ is an important measure of the dimensionality 
of the manifold ensembles appearing in quantum gravity. In particular 
much effort has recently been put into its determination for discretized
two-dimensional quantum gravity ensembles.  The spectral dimension
 is defined by a random walk 
which leaves a fixed vertex at $t=0$ and at every step is allowed to move
from its present position to one of its neighbouring vertices  with uniform
probability. After $t$ steps the probability that the walk has returned to
the initial point is given by
\beq{AA} P(t)\approx \frac{const}{t^{d_s/2}}\eeq	
provided that $t\gg 1$ (to negate discretization effects) and that
$t\ll N^\Delta$ where $N$ is the number of vertices in the graph and
$\Delta$ some exponent (to avoid finite size effects). The spectral dimension
has been investigated numerically for many different values of the central charge $c$ \cite{lots} and it has been calculated analytically for the generic
branched polymer model \cite{me} and found to be $d_s=\fourthirds$ in excellent agreement with the numerical results at large $c$.  We refer the reader to refs
\cite{lots} and \cite{me} for detailed discussion of the definition of 
$d_s$ and the ensemble averages involved. In this letter we extend the results of \cite{me} to the non-generic branched polymers.

We start by recalling the essential features of the  branched
polymer models \cite{MCBP}. A general branched polymer (BP) model has grand canonical partition function $\C Z(z)$ satisfying
\beq{defBP}\C Z(z)=z(1+\sum_{n=1}^\infty\alpha_n \C Z(z)^n)\eeq
where the $\alpha_n$ are constants. 
Iterating this equation shows that $\C Z(z)$ is the generating function
for the set of all rooted  trees $\C B$ made up of links and vertices of all coordination numbers $n+1$ such that $\alpha_n\ne 0$,
\beq{graphical}\C Z(z)=\sum_{{\R A}\in\C B} z^{N_{\R A}} w_{\R A}.\eeq
The number of links in A is denoted by $N_{\R A}$ and the weight $w_{\R A}$
of A is given by
\beq{weight} w_{\R A}=\prod_{v\in \R A}\alpha_{n(v)-1}\eeq
where $v$ runs over all vertices of A and $n(v)$ denotes the coordination
number of $v$. We denote the set of all polymers whose first vertex has coordination number $n+1$ by $\C B^n$; clearly
$\C B =\C B^0\union\C B^1\union\C B^2\union\ldots$ ($\C B^0$ contains
the  polymer consisting of a single link).
 Any graph $\R A\in \C B^n$
 has a set of constituents $\R A_1,\ldots,\R A_n$ obtained by severing the links 
connecting the first vertex to the rest of the graph (see fig.1). 
\begin{figure}[h]
{\epsfxsize=8cm \epsfbox{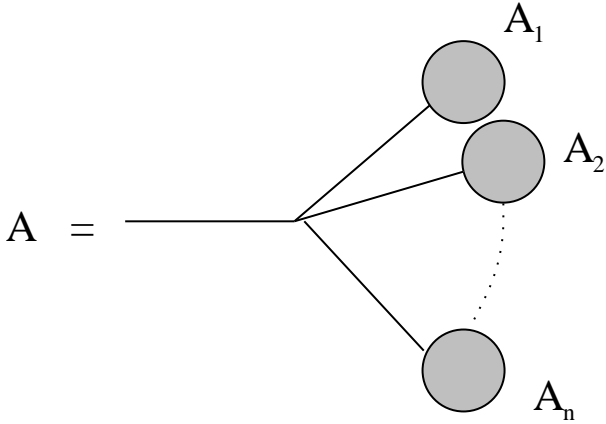}}
\caption{The constituents $\R A_1,\ldots,\R A_n$  of a branched polymer
$\R A\in \C B^n$.}
\end{figure}
 Note that
for $\R A\in \C B^n$
\beq{factor} w_{\R A}=\alpha_n\prod_{i=1}^n w_{{\R A}_i}\eeq
and
\beq{sizes}N_{\R A}=1+\sum_{i=1}^nN_{{\R A}_i}.\eeq

Equation \rf{defBP} can be rewritten in the form
\bea \frac{1}{z}&=&\C F(\C Z)\\
\C F(x)&=&x^{-1}(1+\sum_{n=1}^\infty\alpha_n x^n).\label{2}\eea
For small enough $z$ the solution $\C Z(z)$ is an analytic function but
at some critical value $z=\zcrit$ it is non-analytic; this is the point
 at which
the graphical expansion \rf{graphical} diverges. The susceptibilty is given by
\beq{defsusc}\chi(z)=\frac{\partial\C Z}{\partial z}=
-\frac{z^{-2}}{\C F'(\C Z(z))}=\frac{\C Z z^{-1}}{1-z\sum_{n=1}^\infty n
\alpha_n \C Z(z)^{n-1}}\eeq
and so the critical point is where $\C F'(\C Z(\zcrit))=0$. As $z\to\zcrit$ 
the susceptibility has leading non-analytic behaviour given by
\beq{defgamma}\chi(z)\simeq (\zcrit-z)^{-\gstr}\eeq
where $\gstr$ is the susceptibility exponent.
(Throughout this letter we use the symbol $\simeq$ to denote the leading
singularity as $z\to\zcrit$.) The nature of the
 critical point and the value of $\gstr$ depend upon how  $\C F'$ vanishes. The multi-critical BPs
are obtained by
 supposing that 
$\C F^{(n)}(\C Z(\zcrit))=0,\:n=0,\ldots k$ but that $\C F^{(k+1)}
(\C Z(\zcrit))\ne 0$ so, close to $\zcrit$,
\beq{4}\chi(z)\simeq\frac{z^{-2}}{\vert\C Z(z)-\C Z(\zcrit)\vert^k}\;.\eeq
However
\beq{Zcrit}\C Z(z)\simeq\C Z(\zcrit)-const. (\zcrit-z)^{1-\gstr}\eeq
so comparing singularities we obtain
\bea \gstr&=&k(1-\gstr)\nn\\
{\mathrm {or}}\quad \gstr&=&1-\frac{1}{k+1}\;.\label{6}\eea
The generic case, requiring no special tuning of the $\alpha_n$ (except that
at least one $\alpha_n$ ($n\ge 2$) must be non-zero), is $k=1$,
$\gstr=\half$; higher values of $k$, which require at least $k$ of
the  $\alpha_n$ ($n\ge 2$) to be non-zero and of varying sign to
 enforce the vanishing of higher derivatives, give the multi-critical BPs
(MCBP from now on).

The MCBPs are not the only ones with $\gstr\ne\half$. By allowing an infinite number of the $\alpha_n$ to  be non-zero we can arrange that 
$\C F'$ vanishes non-analytically \cite{Orland,Bialas}.  Consider
\beq{obvious}\C F(\C Z)=\frac{1}{\C Z}+\mu+\frac{\C Z}{A^2}+\lambda
\left(1-\frac{\C Z}{A}\right)^{\beta+1}\eeq
where $A$, $\lambda$, $\mu$ and $\beta$ are positive constants. Then
\beq{obviouser}\C F'(\C Z)=-\frac{1}{\C Z^2}+\frac{1}{A^2}-\frac{\lambda
(\beta+1)}{A}
\left(1-\frac{\C Z}{A}\right)^\beta.\eeq
As $z$ increases from zero so does $\C Z(z)$. The first zero of $\C F'$ is
at $\C Z=A$ when $\zcrit=(2/A+\mu)^{-1}$; provided that $1>\beta>0$ the
non-analytic term in \rf{obviouser} dominates as $z\to\zcrit$. 
 Inserting this dominant behaviour into \rf{4},
 using \rf{Zcrit}, and comparing singularities,
we obtain
\beq{newgamma}\gamma=\frac{\beta}{\beta+1}.\eeq
Thus we get models with continuously varying $\gstr$ in the range 0 to 
$\half$ as  discussed in \cite{Bialas}. 
However the restriction on the range of $\beta$ is easily removed; by
tuning the coefficients $\alpha_n$ so that the analytic part of $\C F'$ in
\rf{obviouser} vanishes as $(\C Z-A)^2$ the non-analytic behaviour dominates for $2>\beta>0$ so now we can get continuously varying $\gstr$s up to 
$\twothirds$.
If  the coefficients are tuned so that the analytic part of $\C F'$ in
\rf{obviouser} vanishes as $(\C Z-A)^m$ then $\beta$ can be as large as $m$.
In this way we see that by combining the multi-critical strategy and the 
non-analytic form of $\C F$ used in \cite{Orland,Bialas}
 we can obtain all $\gstr$ values in the range 0 to 1.
If $\beta$ is an integer these models just reproduce the MCBP; when
$\beta$ is not an integer we will call them the `continuous critical
branched polymers', or CCBP.

The CCBPs actually continue to negative gamma. Suppose that $\C F'$
is finite at $\zcrit$ but that $\C F''$ diverges; then by \rf{defsusc} 
the susceptibility
$\chi$ is finite at $\zcrit$. However
\beq{derivsusc} \frac{\partial \chi}{\partial z}=
\frac{2z^{-3}}{\C F'(\C Z(z))} + \frac{z^{-2}\C F''(\C Z(z))\chi}
{\C F'(\C Z(z)^2) } \eeq
so the derivative of $\chi$ diverges at the critical point and $\gstr$ is
negative. This is easily arranged by (for example) eliminating the linear 
term in \rf{obvious} so that
\beq{neggam}\C F(\C Z)=\frac{1}{\C Z}+\mu+\lambda
\left(1-\frac{\C Z}{A}\right)^{\beta+1}\eeq
with $1>\beta>0$ and comparison of the behaviour as $z\to\zcrit$ then gives
\beq{nng}\gstr=\frac{-\beta}{2-\beta}.\eeq

In our calculation of 
 the spectral dimension we will need to know the behaviour of
\beq{7}\Omega_l=\frac{d^{l+1}}{d\C Z^{l+1}}\sum_{n=1}^\infty 
\alpha_n \C Z(z)^n=\frac{d^{l+1}}{d\C Z^{l+1}} \C Z\C F(\C Z)\eeq
close to the critical point. Using the properties of $\C F$ discussed above
we find that for $\gstr>0$
\bea \Omega_l&\simeq& (\C Z(z)-\C Z(\zcrit))^{\beta-l}+\frac{1}{\zcrit}\delta_{l,0}\nn\\
&=&(\zcrit-z)^{\gstr-l(1-\gstr)}+\frac{1}{\zcrit}\delta_{l,0}\label{omegas}\eea
unless $\beta$ is an integer (ie MCBP) in which case
for $l\ge \beta$ we find that $ \Omega_l$ is finite at $\zcrit$. If 
$\gstr<0$ then $\C F'$ is finite and it follows from \rf{defsusc} that
\beq{spcase} 1-\zcrit\Omega_0(\zcrit)= C\eeq
where $C$ is finite.

Our calculation of the spectral dimension follows the same method as in
\cite{me} but generalized to take account of the presence of vertices of varying order.
The return probability generating function $P_{\R A}(y)$
 on a given polymer $\R A\in\C B^n$ is related to that on its $n$ constituents
$\R A_1,\ldots \R A_n$ (see fig.1). It is convenient to define for
 any polymer
$\R A\in\C B$
\beq{9} P_{\R A}(y)=\frac{1}{1-y}\frac{1}{h_{\R A}(y)}\eeq
and then we find for $\R A\in\C B^n$ \cite{me}
\beq{recur}h_{\R A}(y)=\frac{1+\sum_{i=1}^n h_{{\R A}_i}(y)}{1+(1-y)
\sum_{i=1}^n h_{{\R A}_i}(y)}.\eeq
Note that at $y=1$ the solution is
\beq{easy}h_{\R A}(1)=N_{\R A}\eeq
for all polymers $\R A\in\C B$.
The spectral dimension is found by considering the quantity
\beq{12}\Qt(z,y)=-\frac{d}{dy}(1-y)\sum_{\R A\in\C B} z^{N_{\R A}}w_{\R A}
 P_{\R A}(y)=\frac{d}{dy}\sum_{\R A\in\C B}\frac{ z^{N_{\R A}}w_{\R A}}{h_{\R A}(y)}.\eeq
At $y=1$ the $n$th derivative with respect to $y$ of $\Qt(z,y)$ behaves as
\beq{star} \Qt_n(z)= \frac{\partial^n}{\partial y^n}\Qt(z,y)\yy \simeq (\zcrit-z)^{\beta-n\Delta}\eeq
where the exponents $\beta$ and $\Delta$ are related to the spectral dimension by
\beq{14}\beta=1-\gstr+\Delta(d_s/2-1).\eeq
The behaviour \rf{star} can be established by considering the quantities
\beq{defH} H^{(n_1,n_2,\ldots,n_p)}=\sum_{\R A\in\C B} z^{N_{\R A}}w_{\R A}
\prod_{i=1}^p \frac{d^n}{dy^n}h_{\R A}(y)\yy\eeq
which have leading singular behaviour as $z\to\zcrit$ given by \cite{me}
\beq{singH} H^{(n_1,n_2,\ldots,n_p)}\simeq 
(\zcrit-z)^{a-bp-c\sum_{i=1}^p n_i}\eeq
where we will determine the constants a,b,c.

First we will consider the case when $\gstr>0$. Let us compute
\bea H^{(1)}&=&\sum_{\R A\in\C B} z^{N_{\R A}}w_{\R A}h'_{\R A}(1)\nn\\
&=&\sum_n\sum_{\R A\in\C B^n} z^{N_{\R A}}w_{\R A}h'_{\R A}(1).\eea
Now we use \rf{recur} to relate $h'_{\R A}$ to the corresponding
quantity for the constituents of $\R A$ and \rf{weight} to relate the weights and obtain
\bea H^{(1)}&=&\sum_n\alpha_n\sum_{\R A\in\C B^n}\unionsum\nn\\&&\left\{
\sum_{i=1}^n h'_{{\R A}_i}(1) +\sum_{i=1}^n h_{{\R A}_i}(1)
\left(1+\sum_{j=1}^n h_{{\R A}_j}(1)\right)\right\}.\label{simple}\eea
Note that although $\R A\in\C B^n$ in the above expression there is no such restriction on its constituent polymers, $\R A_i\ldots\R A_n$ which are drawn from
the entire ensemble $\C B$. The terms on the r.h.s. of \rf{simple} which involve $h_{{\R A}_m}'(1)$, of which there are $n$, simply reproduce $H^{(1)}$ multiplied by powers of
the partition function; by using \rf{easy} we see that the remaining pieces simply produce derivatives
of the partition function so we may rearrange \rf{simple} to obtain
\beq{simpler} H^{(1)}\left(1-z\sum_{n=1}^\infty n
\alpha_n \C Z(z)^{n-1}\right)=z^2\frac{\partial^2}{\partial z^2}\C Z.\eeq
The coefficient of $H^{(1)}$ is $1-z\Omega_0$ so by \rf{defgamma} and \rf{omegas} we obtain
\beq{result1}H^{(1)}\simeq (\zcrit-z)^{-1-2\gstr}\eeq
and also, since $H^{(1)}$ is the second derivative of  $\Qt_0(z)$ with respect to $z$,  that 
\beq{15} \Qt_0(z)\simeq (\zcrit-z)^{1-2\gstr}\eeq
which gives $\beta=1-2\gstr$.

Next we compute
\bea H^{(1,1)}&=&\sum_{\R A\in\C B} z^{N_{\R A}}w_{\R A}h'_{\R A}(1)h'_{\R A}(1)\nn\\
&=&\sum_n\alpha_n \sum_{\R A\in\C B^n}\unionsum\left\{\sum_{i=1}^n h'_{{\R A}_i}(1)\right\}^2\nn\\
&&+2z^2\frac{\partial^2}{\partial z^2}\sum_n\alpha_n\sum_{\R A\in\C B^n}\unionsum\left\{\sum_{i=1}^n h'_{{\R A}_i}(1)\right\}\nn\\
&&+\left(z^2\frac{\partial^2}{\partial z^2}\right)^2\C Z .\label{worse}\eea
Again the first sum contains pieces which reproduce $H^{(1,1)}$; the remaining
terms on the r.h.s. of \rf{worse} can be expressed in terms of $H^{(1)}$ and
$\C Z$. We obtain
\bea  H^{(1,1)}\left(1-z\sum_{n=1}^\infty n
\alpha_n \C Z(z)^{n-1}\right)&=&z\sum_{n=2}^\infty n(n-1)\alpha_n \C Z^{n-2}
 \left(H^{(1)}\right)^2\nn\\
&&+2z^2\frac{\partial^2}{\partial z^2} z\sum_{n=1}^\infty n
\alpha_n \C Z(z)^{n-1}H^{(1)}\nn\\
&&+\left(z^2\frac{\partial^2}{\partial z^2}\right)^2\C Z.\label{notsobad}\eea
Noting that the coefficient of the  $\left(H^{(1)}\right)^2$ term  is simply
$z\Omega_1$ and that the coefficient of the $H^{(1)}$ term is $z\Omega_0$ which tends to  1 as $z\to\zcrit$
we see that the first two terms on the r.h.s. of \rf{notsobad} both vary as
$(\zcrit-z)^{-3-2\gstr}$ whilst the last term is sub-leading since it varies as
$(\zcrit-z)^{-3-\gstr}$.
The coefficient of  $H^{(1,1)}$ varies, as before, like $(\zcrit-z)^{\gstr}$
so we find
\beq{hooray}H^{(1,1)}\simeq (\zcrit-z)^{-3-3\gstr}.\eeq
We can iterate this process; $H^{(1,1,1)}$ is given by
\bea  H^{(1,1,1)}\left(1-z\Omega_0\right)&=&z\Omega_2\left(H^{(1)}\right)^3
+z\Omega_1 H^{(1,1)} H^{(1)}+3z^2\frac{\partial^2}{\partial z^2} z\Omega_0 H^{(1,1)}\nn\\
&&+\, \hbox{(less singular terms)}.\label{reallybad}\eea
The presence of the coefficients $\Omega_{1,2}$ in the first two terms on 
the r.h.s. changes their degree of
divergence to $(\zcrit-z)^{-5-3\gstr}$ which is the same as that of the
third term and hence we conclude that
\beq{hoorayagain}H^{(1,1,1)}\simeq (\zcrit-z)^{-5-4\gstr}.\eeq

This process may be continued to longer strings $(1,1,1,\ldots,1)$ and 
to strings involving higher derivatives of $h_{\R A}(y)$ (obtained by
successive differentiation of \rf{recur}). 
 Following \cite{me} we can
set up a proof by induction \cite{joaosthesis} that
\beq{resH} H^{(n_1,n_2,\ldots,n_p)}
\simeq (\zcrit-z)^{1-\gstr-p-(1+\gstr)\sum_{i=1}^p n_i}\eeq
from which it follows, by inserting this result in \rf{star} (see \cite{me}
for details), that
$\Delta=1+\gstr$. Substituting this and $\beta=1-2\gstr$ into \rf{14}
we find that
\beq{killer}d_s=\frac{2}{1+\gstr}.\eeq
Note that the spectral dimension also satisfies the scaling relation
$d_s\Delta=2$. As was discussed in \cite{Zakopane} this scaling relation is expected to be true for graphs whose Laplacian has no level-dependent degeneracy in its eigenvalue spectrum; clearly the BP ensembles, most of whose graphs
have no symmetries, fall into this category.

The case of $\gstr<0$ is simpler. The factor $1-z\Omega_0$ does not vanish 
as $z\to \zcrit$  so we now find that 
\beq{A} H^{(1)}\simeq (\zcrit-z)^{-1-\gstr}\eeq
and therefore that 
\beq{B} \Qt_0(z)\simeq (\zcrit-z)^{1-\gstr}\eeq
which implies $\beta=1-\gstr$.
Using \rf{14} we see immediately that $d_s=2$.  Of course it is necessary
to check that $\Delta\ne 0$ before reaching this conclusion; in fact it is
straightforward to show that $\Delta=1$ \cite{joaosthesis} and so the scaling relation $d_s\Delta=2$ is still satisfied.

It is interesting to compare these results with a scaling relation
recently found by Ambj\o rn et al \cite{janslatest}. They showed that 
\beq{ambjorn} \frac{2}{d_H}=\Delta\left(1-\frac{d_s}{2}\right)\eeq
where $d_H$ is the \emph{extrinsic} Hausdorff dimension. For the $k$th
multicritical model $d_H$ has been calculated \cite{MCBP} and is known
to be given by
\beq{D} d_H=\frac{2(k+1)}{k}.\eeq
Assuming that $d_s\Delta=2$ we can then use \rf{ambjorn} to determine that
\cite{janslatest}
\beq{E} d_s=\frac{2}{2-\frac{1}{k+1}}=\frac{2}{1+\gstr}\eeq
in agreement with our calculations. The value of $d_H$ has not been calculated explicitly for the CCBPs but we can now use the scaling relation \rf{ambjorn} and our results  to show that $\gstr d_H=2$ if $\gstr>0$.  When $\gstr<0$
we have $d_s=2$ so we expect $d_H=\infty$.  It is amusing that 
the two-dimensional quantum gravity models at $c<1$ also have $d_H=\infty$
and hence $d_s=2$ \cite{janslatest}.

\vspace{1 truecm}
\noindent We acknowledge valuable conversations with Jan Ambj\o rn who
told us of the scaling relation result \rf{ambjorn} prior to writing the
paper \cite{janslatest}.
J.C. acknowledges  a grant from {\sc Praxis XXI}.

\end{document}